\documentclass[preprint,prb,aps,showpacs,footinbib,amsmath,amssymb,groupedaddress]{revtex4-1}
\usepackage{graphicx}
\usepackage{dcolumn}
\usepackage{color}
\usepackage{bm}
\usepackage{natbib,hyperref}
\usepackage{amsmath,amssymb,amsfonts,bbold}
\usepackage[normalem]{ulem}

\newcommand{\beq}{\begin{equation}}
\newcommand{\eeq}{\end{equation}}
\newcommand{\beqa}{\begin{eqnarray}}
\newcommand{\eeqa}{\end{eqnarray}}

\begin{document}


\title{Breaking crystalline symmetry of epitaxial SnTe films by strain}



\author{Steffen Schreyeck}
\email{sschreyeck@physik.uni-wuerzburg.de}
\author{Karl Brunner}
\author{Laurens W. Molenkamp}
\affiliation{Institute for Topological Insulators and Physikalisches Institut, Experimentelle Physik III, 
	Universit\"{a}t W\"{u}rzburg, Am Hubland, 97074 W\"{u}rzburg, Germany}
\author{Grzegorz Karczewski}
\affiliation{Institute of Physics, Polish Academy of Sciences, Aleja Lotnik$\acute{o}$w 32/46, 02-668 Warsaw, Poland}

\author{Martin Schmitt}
\email{maschmitt@physik.uni-wuerzburg.de} 

\author{Paolo Sessi}

\author{Matthias Vogt}

\author{Stefan Wilfert}

\author{Artem~B.~Odobesko}

\author{Matthias Bode}

	\affiliation{Physikalisches Institut, Experimentelle Physik II, 
	Universit\"{a}t W\"{u}rzburg, Am Hubland, 97074 W\"{u}rzburg, Germany}

\date{\today}

\begin{abstract}
SnTe belongs to the recently discovered class of topological crystalline insulators. 
Here we study the formation of line defects which break crystalline symmetry by strain in thin SnTe films. 
Strained SnTe(111) films are grown by molecular beam epitaxy 
on lattice- and thermal expansion coefficient-mismatched CdTe. 
To analyze the structural properties of the SnTe films 
we applied {\em in-situ} reflection high energy electron diffraction, x-ray reflectometry, 
high resolution x-ray diffraction, reciprocal space mapping, and scanning tunneling microscopy. 
This comprehensive analytical approach reveals a twinned structure, 
tensile strain, bilayer surface steps and dislocation line defects forming a highly ordered dislocation network for thick films with local strains up to 31\% breaking the translational crystal symmetry. 

\end{abstract}


\maketitle 

\section{Introduction}

The recently discovered new material class of topological insulators attracted a lot of interest in condensed matter physics.\cite{Kane2005,Bernevig2006,Konig2007} Fu extended the topological classification of band structures by including crystal point group symmetries, i.e., the theoretical prediction of topological crystalline insulators (TCIs).\cite{Fu2011} In contrast to topological insulator materials, where the surface states are protected by time reversal symmetry, the surface states of TCIs are protected by the crystal symmetry, resulting in an insulating bulk and metallic surface states on high symmetry crystal surfaces, such as (001) and (111) planes.\cite{Xu2012} The first experimentally discovered TCIs belong to the cubic rocksalt family, namely the lead-tin salts Pb$_{1-x}$Sn$_{x}$Se for $x\geq0.2$, SnTe, and Pb$_{1-x}$Sn$_{x}$Te for $x\geq0.4$. \cite{Dziawa2012,Hsieh2012,Tanaka2012} For the latter the spin-polarized nature of the Dirac surface states has been observed by spin-resolved ARPES measurements.\cite{Tanaka2012} The reduction of crystal symmetry opens the possibility to establish energy shifts or band gaps.\cite{Barone2013, Zhao2015} Sessi {\em et al.} showed that breaking the translational symmetry at odd surface steps on cleaved Pb$_{1-x}$Sn$_{x}$Se(001) bulk crystals results in topological 1\,D edge states. \cite{Sessi2016} Such line defects may allow for the creation of well-separated conductive channels, which can be patterned and contacted in thin TCI films for spintronics devices.\cite{Sessi2016}

We investigate molecular beam epitaxy of rocksalt TCI thin films to control the crystal orientation and the formation of crystalline line defects at the surface. The IV-VI rocksalt structure is known to build a glide plane system under epitaxial and thermal strains forming dislocations at the interface to the substrate, which extend to the surface.\cite{Bauer1986,Zogg1995,Springholz2007} The MBE growth of SnTe has been studied on various substrates, including Si, BaF$_2$, Bi$_2$Te$_3$, PbSe, and CdTe \cite{Yan2014,Akiyama2016,Taskin2014,Zeljkovic2015,Ishikawa2016}. Here we study epitaxy of thin SnTe(111) films, a facet that is hardly accessible by crystal cleaving,\cite{Takana2013} on lattice mismatched CdTe(111) to obtain in-plane tensile strain. For strain analysis we grow films in the thickness range from 8.5\,nm to 425\,nm on CdTe. SnTe crystallizes in rocksalt structure with a room temperature lattice mismatch of 2.6\% to the cubic zincblende lattice of CdTe. In addition the linear thermal expansion coefficient of SnTe ($2\times10^{-5}$/K) is about a factor of three larger than that of CdTe ($6\times10^{-6}$/K) introducing additional in-plane tensile strain at low temperatures.\cite{Khokhlov2002} To study the structural quality and strain in the layers, we apply {\em in-situ} reflection high energy electron diffraction (RHEED), x-ray reflectometry (XRR), high resolution x-ray diffraction (HRXRD), and HRXRD reciprocal space mapping. The structural properties of the pristine SnTe(111) surface is compared to the dislocation network by means of scanning tunneling microscopy (STM) at low temperatures.

\section{Experimental}

The SnTe films are grown by molecular beam epitaxy under ultra high vacuum (UHV) conditions 
at a base pressure below 1$\times$ 10$^{-10}$ \,mbar. 
As a virtual substrate we use a several~$\mu$m thick CdTe buffer layer, 
which is grown on a two inch Si-doped n-type GaAs(111)B wafer (B implies As-terminated surface). 
In between the buffer and the wafer a 6\,nm thin ZnTe film is deposited to reduce the lattice mismatch of about 13\% 
and to enable CdTe(111) growth.\cite{Grynberg1996} 
After the buffer growth this substrate is divided into about 1\,cm $\times$ 0.5\,cm sized pieces suitable for STM experiments. 
This {\em ex-situ} procedure assures that all future experiments 
will be performed on substrates with comparable quality.
The individual substrate pieces are etched by 12\% aqueous HCl solution for 30~seconds to remove the oxide layer 
before they are indium-glued on an UHV sample holder. 
The holders are equipped with a moveable tantalum spring for providing electrical top contact 
to the sample surface after growth allowing further STM characterization.

\begin{figure*}[tbh]   
\centering
\includegraphics[width=1\linewidth]{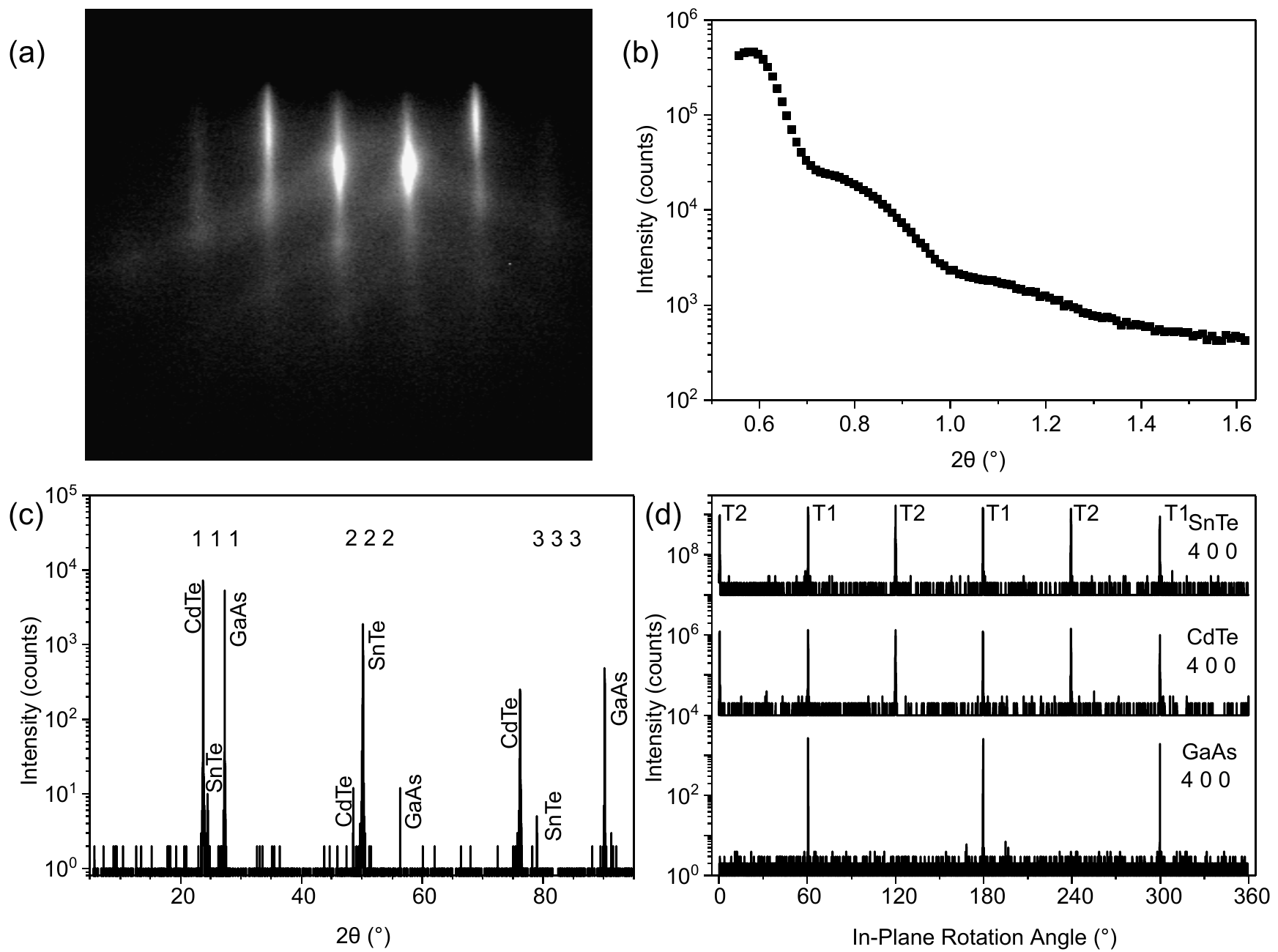}%
\caption{(a) Representative RHEED image taken after 170\,nm SnTe MBE growth. 
	(b) XRR measurement of a thin SnTe film to determine the SnTe layer thickness, i.e., 17\,nm. 
	(c) Large overview HRXRD $\theta$-2$\theta$ diffractogram of the 425\,nm thick SnTe sample 
	displaying the 1\,1\,1 , 2\,2\,2, and 3\,3\,3 reflections of GaAs, CdTe, and SnTe. 
	(d) HRXRD pole scans of the asymmetric 4\,0\,0 reflections of 425\,nm thick SnTe sample. 
	The two triplets of twin reflections are labeled T1 and T2. } 
\label{fig1}
\end{figure*}

After reintroduction into the MBE chamber, 
the RHEED pattern of the etched CdTe surface shows a mixture of amorphous and polycrystalline features, 
which is caused by nearly unordered elemental Te remaining on the CdTe buffer layer after {\em ex-situ} HCl etching. 
To restore the surface quality the substrates are first heated 
to the CdTe buffer growth temperature of 300$^\circ$C, 
thereby desorbing excess Te, as confirmed by a change of the RHEED pattern from diffuse and spotty to a streaky pattern. 
After Te desorption, the bare CdTe surface is protected by applying Te flux and  
 an additional 0.5~$\mu$m CdTe layer is grown on the CdTe (111) buffer to renew the surface. 
Most part of this layer is grown at 300$^{\circ}$C, but
the sample is cooled to the SnTe growth temperature of 260$^{\circ}$C towards the end of CdTe growth, when a Cd rich surface is prepared by applying Cd flux. 
Immediately after closing the Cd shutter, the SnTe growth is initiated by co-deposition of Sn and Te. 
The applied elemental fluxes of Sn and Te have constant beam equivalent pressures 
of 9\,$\times$\,10$^{-7}$\,mbar and  9\,$\times$\,10$^{-6}$\,mbar, respectively. 
Beam equivalent pressures are measured by an ionization gauge (after Bayard-Alpert) 
and are corrected by ionization sensitivity, source temperature, and mass of the impinging particles. 
The structural properties of the SnTe films are analyzed by {\em in-situ} RHEED 
and a Panalytical Xper't MRD high resolution x-ray diffractometer (HRXRD) equipped with Cu K$_{a1}$ source.
The SnTe layers investigated in this study were grown for 250\,s, 500\,s, 2500\,s, and 12500\,s 
to obtain a series with different film thicknesses.

After the growth a tantalum spring is positioned {\em in-situ} on top of the SnTe film 
by a wobble stick to ensure a proper electrical grounding for STM measurements. 
Afterwards the samples are transferred into an UHV-suite case 
and transported to the STM setup at a base pressure below $5 \times 10^{-10}$ \,mbar.
Scanning probe experiments are carried out in a home-built low-temperature 
scanning tunneling microscope operating at a temperature of $T\approx 5.5$ \,K. 
All topographic images are obtained in constant-current mode with tungsten tips.

\section{Results and discussion}

\subsection{Epitaxy}
\label{sec:CrystStruc}
The CdTe buffer growth on the GaAs(111)B substrate is monitored by RHEED. 
While the streaky RHEED pattern of the {\em in-situ} grown CdTe indicates a 2D growth mode, 
the second layer after {\em ex-situ} etching shows streaks and spots (data not shown here). 
Whereas the sharp streaks confirm the presence of atomically flat areas on the CdTe buffer, 
the spots result from a partially rough surface after {\em ex-situ} preparation. 
The symmetrical alignment of the spots indicates a twinned CdTe layer structure.
Subsequently the growth of the SnTe on the CdTe buffer results 
in a decreasing intensity of the spots and increasing intensity of the streaks in the RHEED pattern. 
After a few nanometers of SnTe the spots vanish and a streaky RHEED pattern of a ($1\times1$) ordered hexagonal surface with distinct Kikuchi-lines remains 
indicating a predominant 2D growth-mode at the surface, as shown in Fig.~\ref{fig1}(a). 

\subsection{Bulk crystal structure and strain}
XRR measurements are conducted to determine the layer thickness of the SnTe films. 
Figure~\ref{fig1}(b) shows a XRR curve with distinct fringes resulting in a SnTe film thickness of $17\pm 2$\,nm. 
The layer thicknesses of the SnTe series are 8.5\,nm, 17\,nm , 85\,nm and 425\,nm, 
as extrapolated from the XRR thickness measurement versus the growth time, 
resulting in a growth rate of 0.34\,{\AA}s$^{-1}$.

The large scale $\theta$-$2\theta$ HRXRD measurements reveal exclusively 
the 1\,1\,1 , 2\,2\,2 , and 3\,3\,3 reflections of GaAs, CdTe, and SnTe, as shown in Fig.~\ref{fig1}(c), 
confirming the parallel alignment of the GaAs, CdTe, and SnTe (111) surface planes, 
as well as indicating the absence of other crystalline phases and orientations. 
Pole scans in Fig.~\ref{fig1}(d) of the asymmetric reflections of the $\left\{400\right\}$ planes 
display the expected three-fold symmetry of the GaAs(111) substrate. 
In contrast, the CdTe buffer and the SnTe reflections show two peak triplets resulting in a six-fold symmetry, 
thereby confirming the presence of twinned domains rotated by $180^{\circ}$ to each other. 
One of the peak triplets, labeled T1 in Fig.~\ref{fig1}(c), 
is located at the same in-plane rotation angle as the GaAs and CdTe reflections 
confirming parallel orientation of  the $\left\{400\right\}$ planes of twin T1 with that of the substrate and the buffer. 
This indicates a well aligned crystalline interface between the rocksalt and zincblende crystal structures. 
The comparable peak intensities of triplets T1 and T2 suggest a nearly equal twin distribution for SnTe, which also applies for the CdTe buffer. 
The formation of twin domains is a well known crystal defect in epitaxial CdTe on (111) substrates 
and is attributed to the presence of domains with different stacking orders, i.e., ABC and ACB.\cite{Chen1995}

\begin{figure*}[tb]   
\centering
\includegraphics[width=1\linewidth]{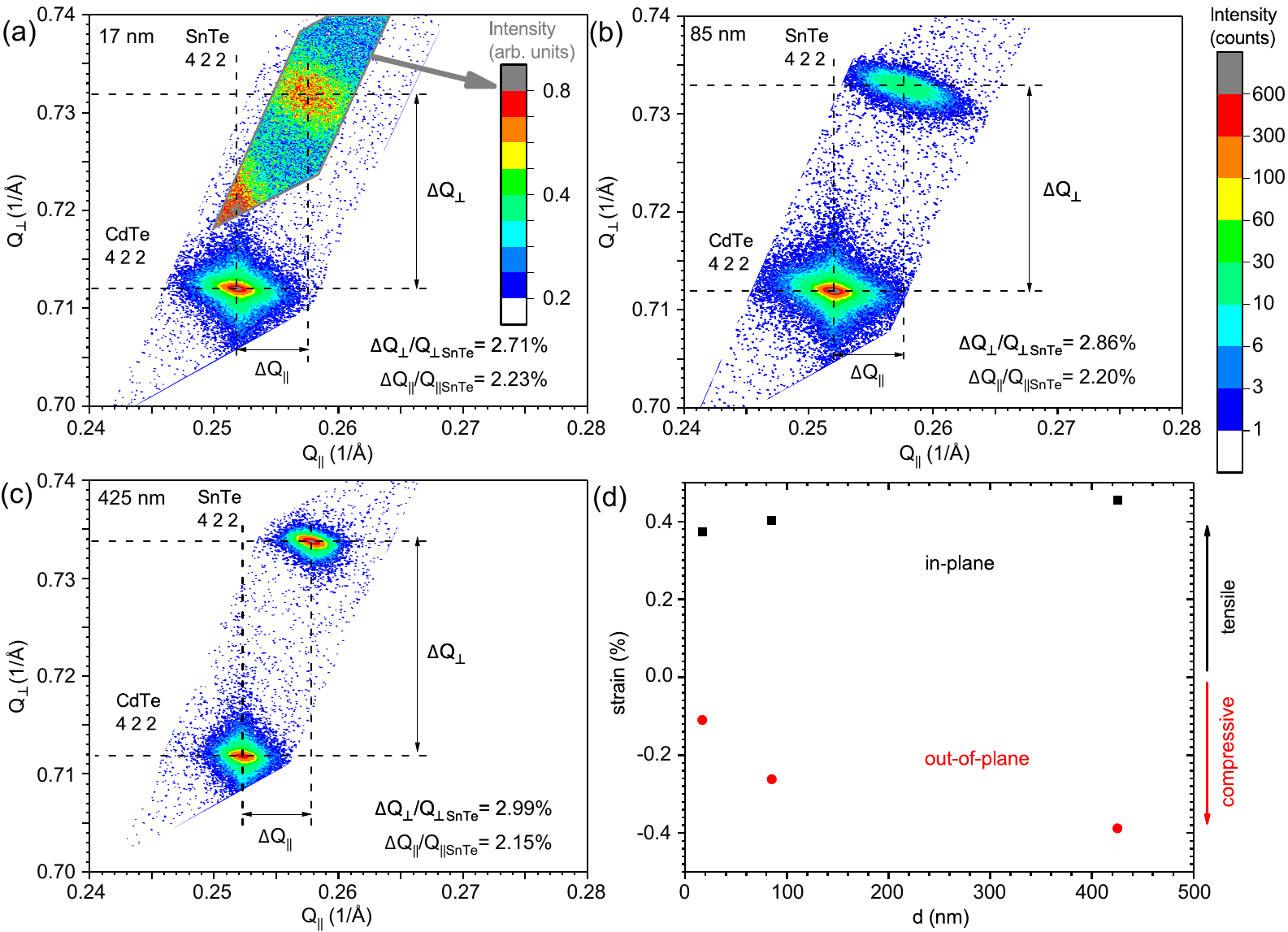}%
\caption{Reciprocal space maps of the CdTe and SnTe 4\,2\,2 reflections of samples 
	with (a) 17\,nm (b) 85\,nm, and (c) 425\,nm with the intensity scale in the upper right part. 
	Note the inset in the grey frame in (a) of the SnTe reflection is measured 
	with 30 times higher integration time and has therefore an other intensity scale. 
	(d) Strain of SnTe determined by the relative deviation from the CdTe 4\,2\,2 reflection position 
	in dependence of the film thickness.} 
\label{fig2}
\end{figure*}
HRXRD reciprocal space maps (RSM) are used to study the strain within the layers at room temperature, 
after cooling the samples from growth temperature to 5.5\,K. 
Note that the thermal treatment of the sample, which is an unavoidable need towards our STM investigation, 
significantly impacts the formation of defects and strain, as will be discussed later in Section~\ref{sec:linedefects}. 
The 4\,2\,2 reflections of buffer and film, which are both accessible in $\omega$-plus and $\omega$-minus geometry 
($2\theta$-$\omega$ diffractogram with $\omega=\theta \pm \angle ([111],[422]))$, are investigated. 
The angle between the $\left[111\right]$ and $\left[422\right]$ direction 
is 19.47$^{\circ}$ in a cubic relaxed zincblende or rocksalt structure. 
The RSMs of the 17\,nm, 85\,nm and 425\,nm films are plotted in Fig.~\ref{fig2}(a)-(c). 
They show an intense 4\,2\,2 reflection of the CdTe buffer 
appearing under an angle of 19.49$^{\circ}$ to the $\left[111\right]$ direction, 
slightly above the expected value, indicating that the CdTe is nearly completely relaxed. 
With increasing layer thickness the intensity of the SnTe 4\,2\,2 reflections increases 
and their width in $\omega$ and $2\theta$ decreases. 
The angle between the $\left[111\right]$ direction and the measured 4\,2\,2 reflection of SnTe, 
$\alpha=19.36^{\circ}$ for the 425\,nm film, is significantly below the calculated value in a relaxed crystal, 
indicating a distortion of the cubic rocksalt structure. 

This distortion is analyzed in more detail by comparing the in-plane and out-of-plane lattice constants 
of the strained SnTe relative to that of the relaxed CdTe buffer. 
Therefore the difference of the Q$_\parallel$ and $Q_\bot$ positions of the 4\,2\,2 reflections are determined. 
The relative difference between lattice constants of the CdTe buffer and the SnTe layers 
are 2.23\%, 2.20\%, 2.15\% within the film plane and 2.71\%, 2.86\%, and 2.99\% normal to the film plane for the 17\,nm, 85\,nm, and 425\,nm films, respectively. 
The deviation from the expected literature value of 2.6\% mismatch is plotted in Fig.~\ref{fig2}(d). 
The in-plane strain is tensile and the out-of-plane strain is compressive, 
as expected for the growth on the larger lattice of CdTe. 
Surprisingly, these values indicate an increase of strain with increasing layer thickness. 
We assume the unexpected increase of strain with thickness 
is either related to an enhanced relaxation of thin films due to island growth or the thermal expansion coefficient mismatch between SnTe and CdTe of about a factor of three. 
The thermal mismatch combined with the thermal treatment of the samples before the HRXRD analysis, 
i.e., the cooling from 260$^{\circ}$C (533\,K) growth temperature 
to 5.5\,K for STM measurements and successive warming to ambient conditions, 
may cause this uncommon strain behavior in these films.\cite{Zogg1995}

\subsection{Topography and surface structure}
To investigate the structural properties of the SnTe(111) surface 
and the consequences of the in-plane strain we performed STM measurements 
on all four samples with film thicknesses ranging from 8.5\,nm up to 425\,nm at low temperatures T=5.5\,K. 
In the case of the 8.5\,nm film was found to be electrically insulating, making it inaccessible by STM. 
We found two plausible explanations for this behavior.
First, a hybridization of surface states located at the top and bottom of the film 
could open a band gap resulting in an insulating SnTe film. 
Second, an inhomogeneous coverage of the substrate could lead to a discontinuous film
with insufficient electrical grounding through the tantalum spring.

\begin{figure*}[tb]   
\centering
\includegraphics[width=1\linewidth]{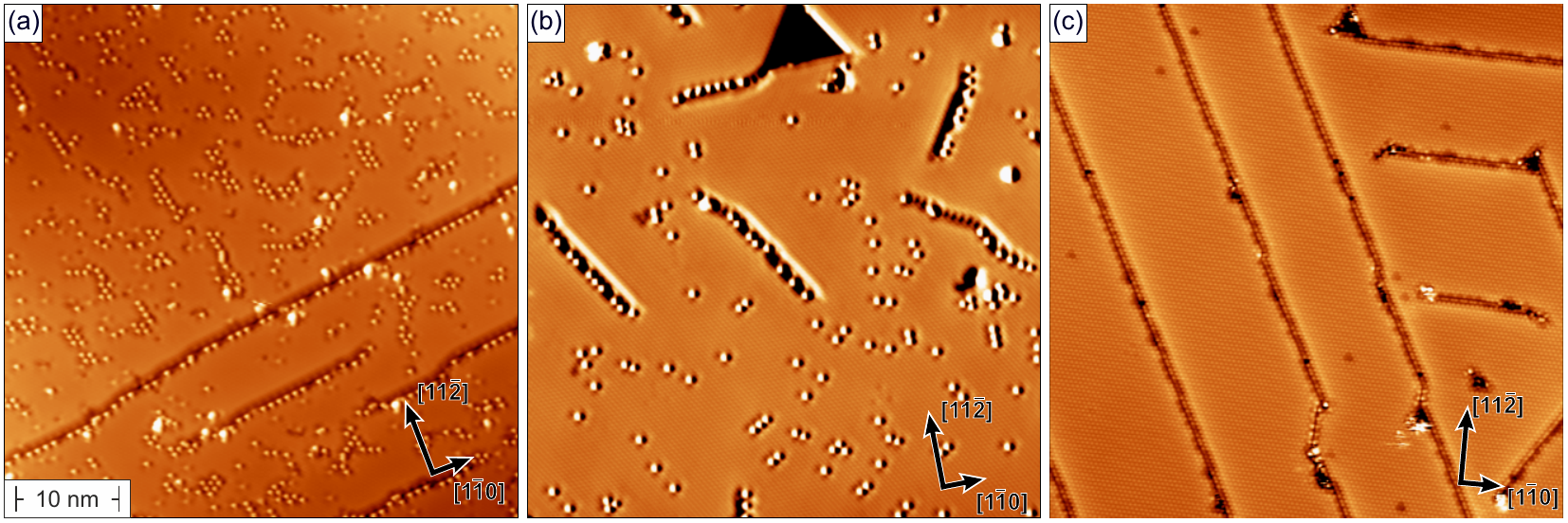}%
	\caption{Representative STM topographic images (50\,nm $\times$ 50\,nm) 
		of SnTe(111) films with a thickness of (a) 17\,nm, (b) 85\,nm, and (c) 425\,nm. 
		In each case an area without step edges but occasional adatoms (white dots) 
		and dislocation lines (dark lines) is shown.  
		While the adatom density decreases with film thickness, 
		the number of dislocation lines appears to increase and even form a dislocation network. 
		Scan parameters: (a) $U = 1$\,V, $I = 300$\,pA; (b) $U = 0.5$\,V, $I = 500$\,pA; 
			(c) $U = 0.8$\,V, $I = 300$\,pA.} 
\label{Fig:thickness}
\end{figure*}
As shown in Fig.~\ref{Fig:thickness} the thicker SnTe(111) films could be imaged by STM. 
The topography of the 17\,nm thick film, as depicted 
in Fig.~\ref{Fig:thickness}(a), is atomically flat and slightly buckled. 
Besides local point-like defects we can also recognize dark lines which are oriented 
along $\langle1\overline{1}0\rangle$ directions and probably represent distinct dislocation lines. 
We also observe these dislocation lines for the 85\,nm film in Fig.~\ref{Fig:thickness}(b) 
and---most strikingly---they become even more abundant 
for the 425\,nm film shown in Fig.~\ref{Fig:thickness}(c)
where periodic dislocation lines along $\langle1\overline{1}0\rangle$ show up. 
This finding indicates  that strain and relaxation process of SnTe films grown on CdTe 
comes along with a thickness-dependent dislocation line network, which may increase strain with increasing layer thickness during thermal treatment, as
observed by HRXRD above in Sect.\,\ref{sec:CrystStruc}.

In the following we will focus on 425\,nm thick films which exhibit periodic dislocation lines 
with a length of up to several hundreds of nanometer and occasional, well separated atomic scale defects. 
Our investigations revealed that the structural properties of both, 
point and line defects, of these rather thick films are representative also for thinner SnTe films on CdTe, 
with the only difference being the density of the particular defects. 

\begin{figure*}[t]   
\centering
\includegraphics[width=1\linewidth]{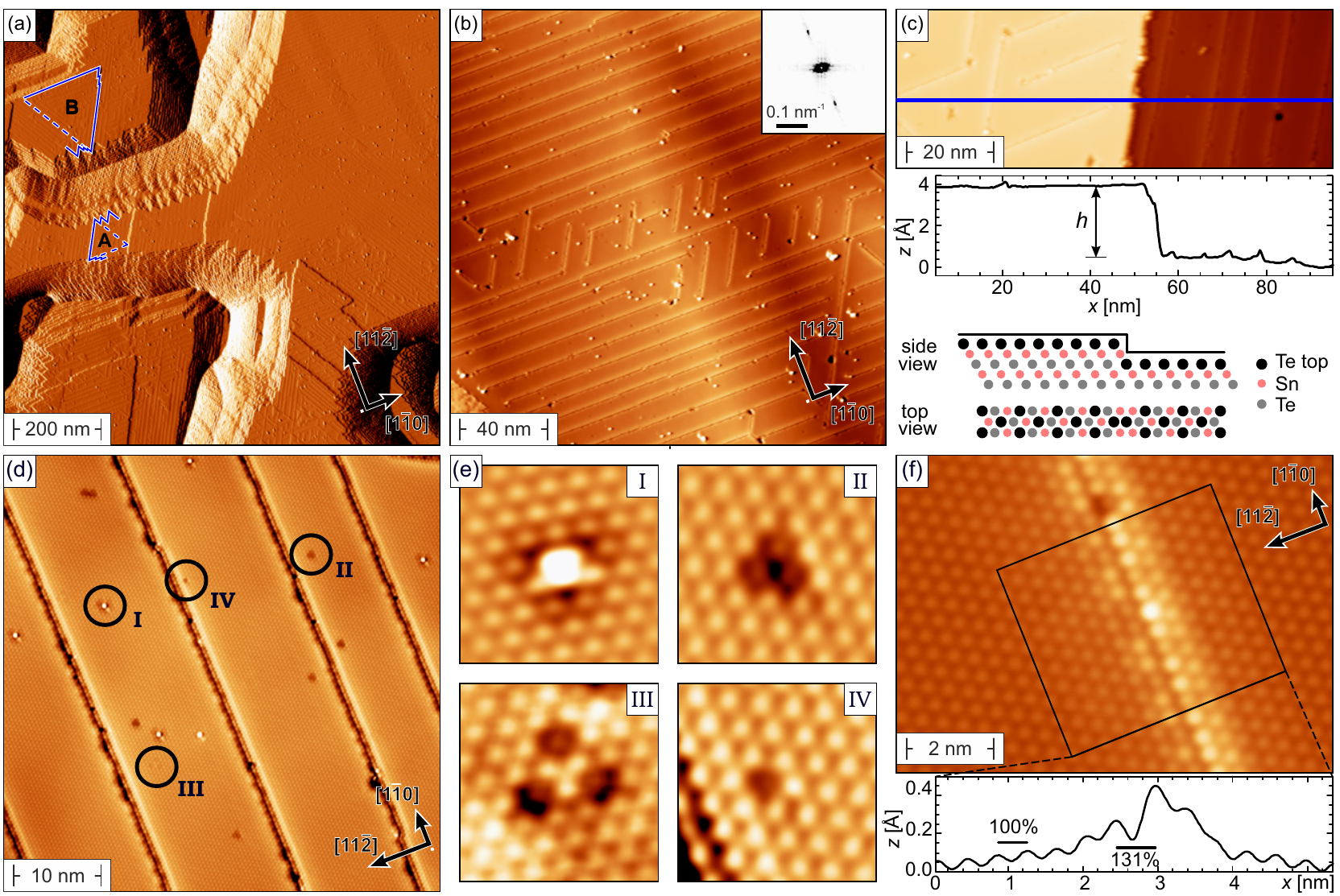}%
	\caption{(a) Overview STM scan of the 425\,nm thick SnTe(111) film on CdTe.  
		Mesas with a height of several nm and triangular shaped edges, indicated in blue, can be recognized. 
		(b) Higher resolution scan of the dislocation network on an extended area without step edges 
		but numerous periodically arranged dislocation lines oriented along $\langle 1\overline{1}0\rangle$ directions. 
		Inset:  Fourier-transformed of the image in (b).   
		Analysis of the spot separation results in a dislocation line periodicity of $(8 \pm 1)$\,nm. 
		(c) STM image of a surface area with a step edge.  
		The line profile shows a step height $h = (3.75 \pm 25)$\,{\AA}. 
		The sketch shows in the side view and the top view the broken translation symmetry at step edges.  
		(d) Large scale atomic resolution scan showing four typical surface defects (I-IV).
		(e) Close-ups of defects I-IV marked in (d). 
		(f) Averaged height profile of atomic resolution data perpendicular to a dislocation line (black rectangle) indicates lateral translation of atomic rows perpendicular to the dislocation by 31\%.
		Scan parameters: (a)-(c) $U = 2$\,V, $I = 50$\,pA; 
			 (d),(e) $U = -0.8$\,V, $I = 300$\,pA; (f) $U = +0.8$\,V, $I = 400$\,pA.} 
\label{Fig:overview}
\end{figure*}
The large overview STM scan of a 425\,nm film in Fig.~\ref{Fig:overview}(a) 
is typical for those films and comparable to the work of Ishikawa {\em et al.}, 
who optimized the growth parameters of SnTe on CdTe substrates.\cite{Ishikawa2016} 
The surface is covered by large mesas, i.e., relatively flat areas often exhibiting extended atomically flat terraces, 
which are separated by steep cliffs forming up to 15\,nm deep trenches. 
Close inspection shows that these cliffs are not abrupt but consist 
of a dense sequence of steps with step heights that are equivalent to single bilayers. 
As sketched by blue triangles in Fig.~\ref{Fig:overview}(a) the cliffs surrounding the mesas 
as well as the few step edges on their top are oriented along $\langle 1\overline{1}0\rangle$ directions, 
whereby the orientation of the triangles rotate by 180$^{\circ}$ between neighboring mesas.
This result is consistent with the existence of twin domains with different stacking order,\cite{tarakina2014} 
as already obtained from the HRXRD pole scans displayed in Fig.~\ref{fig1}(d). 

In addition to occasional step edges oriented in $\langle1\overline{1}0\rangle$ directions, 
we observe wide areas with parallel, periodically arranged dislocation lines 
on top of the mesas, as depicted in Fig.~\ref{Fig:overview}(b). 
These lines are directed in the very same high-symmetry directions as the step edges.  
Their mean separation can be obtained from the Fourier transformation of the STM image, 
which is displayed as an inset in the top right of Fig.~\ref{Fig:overview}(b).  
An analysis of the separation of the two outer spots from the central spot 
located at zero reciprocal length results in a periodicity of $(8 \pm 1)$\,nm. 
Furthermore, we can recognize dark and bright areas with ridges along the [$11\overline{2}$] direction 
which are caused by a surface which is buckled on a lateral scale of about 100\,nm, 
also indicating the presence of strain in the film.
		
Figure~\ref{Fig:overview}(c) shows an STM image of a surface area with a step edge. 
The height profile measured along the blue line is shown in the central panel.
In the lower panel a sketch of the side- and top view is sketched. 
Due to the fcc stacking a natural breaking of the translational symmetry exists, 
which could potentially lead to a topological edge state, similar to what has been observed 
for Pb$_x$Sn$_{(1-x)}$Se(001) surfaces obtained by cleaving.\cite{Sessi2016}
The step height amounts to $h = (3.75\pm0.25)$\,{\AA}, corresponding well to the literature value 
for a bilayer step edge between two Te (Sn) layers of 3.64\,{\AA}. 
This indicates that the (111) surface is either exclusively Te- or Sn-terminated. 
Otherwise, step edges with a height of half this value would be expected. 
Such step edges which would correspond to a transition 
from a Sn- to a Te-terminated terrace (or {\em vice versa}) have never been observed in these films. 
Theoretical investigations of the stability of the SnTe(111) surface 
predict three energetically stable surface structures.\cite{Wang2014} 
The $(1\,\times\,1)$-ordered Te surface for Te-rich growth, 
a $(\sqrt{3}\,\times\,\sqrt{3})$R30$^\circ$-reconstructed surface at higher tin partial pressure, 
and a $(2\,\times\,1)$ Sn reconstruction for tin-rich growth. 
Based on the fact that we grow our samples in Te-rich conditions, 
together with the theoretical predictions of the stability of the SnTe(111) surface and the observed ($1\times1$) streaks by RHEED, we conclude that we see a Te-terminated surface.  

This assumption is also supported by the higher resolution STM scan shown in Fig.~\ref{Fig:overview}(d). 
We see five dislocation lines oriented along the [$1\overline{1}0$] direction. 
In between these line defects we observed a hexagonal lattice with a lattice constant of $(4.72 \pm 0.30)$\,{\AA}. 
This is in agreement with a $(1\,\times\,1)$-ordered surface, as determined by RHEED. 
Besides that we only observe very few point defects, 
the most frequent types of which are exemplarily marked by black circles. 
Higher resolution scans of these defects are shown in Fig.~\ref{Fig:overview}(e). 
They have a typical appearance similar to other TI surfaces 
and can be found at all studied film thicknesses.\cite{Jiang2012} 
Defect (I) to (III) have the defect center in a three-fold coordinated hollow site between the atoms of the top Te layer. 
Since defect (I) exhibits an about 0.81\,{\AA} high protrusion in its center, we assign it to an adatom, probably Te.  
In contrast, the corrugation of defects (II) and (III) is much smaller, 
probably because these defects are located in the first and deeper subsurface layers, respectively. 
Defect IV appears as a 0.25\,{\AA} deep depression centered at a position of the surface lattice.  
Therefore, we ascribed it to a Te vacancy in the surface layer.

\subsection{Discussion of line defects}
\label{sec:linedefects}
 
In the following we want to discuss possible scenarios for the formation of the dislocation line defects.
First, they could act as domain boundaries between regions with different stacking sites, e.g. A and B. 
This would cause an abrupt lateral shift of the atomic rows of 2.6\,{\AA}
by crossing the defect line in Fig.~\ref{Fig:overview}(f). Here the atoms in close proximity to the dislocation are displaced perpendicular to the line defect by 31\%, i.e, 1.2\,{\AA}, 
probably to cope with the misfit strain between SnTe and CdTe, as indicated by the line profile across the line defect.
A closer inspection of the overview scans in Fig.~\ref{Fig:overview}(b) and (c) 
reveals that the dislocation lines are not infinitely long but terminate at specific locations and merge to one domain with the same stacking positions.   
The second scenario is a periodic dislocation patterning process 
in which lattice planes 
are added to or removed from the ends of dislocations. 
However, in contrast to compressively strained rocksalt (001) films 
we observe no additional or missing planes at the ends of dislocation lines.\cite{Springholz2001}
Third, the rocksalt (111) films are known to form dislocations under compressive strain, 
with their threading ends moving back and forth in 
$\left\langle 1\overline{1}0 \right\rangle$ surface directions 
along $\left\{001\right\}$ planes of the bulk, thereby forming
a $\left\langle 1\overline{1}0 \right\rangle$ $\left\{001\right\}$ glide system during thermal cycling.\cite{Zogg1995}
We also observe dislocation lines oriented 
in $\left\langle 1\overline{1}0 \right\rangle$ directions. 
In contrast to Zogg {\em et al.},\cite{Zogg1995} the in-plane strain 
in our system is tensile and the lattice mismatch is lower by a factor of three. 
Therefore, we observe a locally increased lattice constant of up to 31\% breaking the translational symmetry, as shown in Fig.~\ref{Fig:overview}(f), perpendicular to the dislocation lines 
due to the tensile strain introduced by the CdTe. Whereas, first-principle calculations on SnTe indicate a closing of the gap and a topological phase transition to trivial band order already at 3\% of tensile strain. \cite{Hsieh2012} This suggests that a topological phase transition is likely to take place at the dislocation lines observed by STM. 
The mean increase of the in-plane lattice, measured perpendicular to dislocation lines, is $0.82 \pm 0.14$\,{\AA } 
distributed over several atomic positions. 
This corresponds to a mean elastic strain relaxation of 1.0\%  in $\left\langle 11\overline{2} \right\rangle$ direction for the observed 
averaged periodical separation of dislocations, i.e., 8\,nm, at a temperature of 5.5\,K.  The observed mean strain relaxation of about 1.0\% is comparable to the expected thermal mismatch formed during cooling the sample from growth temperature to 5.5\,K.\cite{Knox2014} Therefore SnTe appears to grow nearly relaxed forming a misfit dislocation network of 8 nm period at the interface to the CdTe buffer to overcome the lattice mismatch and the strain measured by XRD is build up due to thermal expansion coefficient mismatch between layer and buffer. 
The thermal cycling of the thin films with low dislocation density 
results in threading ends moving back and forth to relax the thermal strain.\cite{Zogg1995}
Whereas the films with higher dislocation densities and layer thicknesses 
have an increased probability of interaction and pinning of dislocations resulting in higher residual strain, 
as shown by HRXRD reciprocal space maps at room temperature.\cite{Zogg1995} 
Therefore we expect the tensile strain relaxation at glide planes to be the most likely explanation for the line defect formation, 
since it can also describe the uncommon strain behavior in our films.

\section{Conclusion}
In summary, we present a study of strained topological crystalline insulator SnTe(111) thin films grown by MBE on CdTe, 
a surface which is  hardly accessible by {\em in-situ} crystal cleaving. 
The structural analysis by HRXRD reciprocal space maps of these films 
give insight into an uncommon strain relaxation mechanism resulting in residual strain increasing with layer thickness. 
We address this behavior to the formation and pinning 
of dislocations during the cooling procedure after growth due to the thermal expansion coefficients mismatch. 
STM measurements reveal bilayer surface steps and a network of highly ordered dislocation lines 
on the surface along the $\left\langle 1\overline{1}0 \right\rangle$ directions. 
Local strains up to 31\% breaking the translational crystal symmetry are measured perpendicular to the several hundreds of nanometers long dislocation lines.
These compensate the thermal expansion and lattice mismatch between SnTe and CdTe 
and form together with surface steps a playground for the study 
of topological surface states of this topological crystalline insulator with broken translational symmetry.

\begin{acknowledgments}
 We gratefully acknowledge the financial support of the National Science Centre (Poland) through grant UMO-2017/25/B/ST3/02966, and the DFG through SFB
1170 “ToCoTronics” (projects A02 and A05) and the Leibniz Program.\\

S.S. and M.S. contributed equally to this work.
\end{acknowledgments}


\begin{thebibliography}{28}%
\makeatletter
\providecommand \@ifxundefined [1]{%
 \@ifx{#1\undefined}
}%
\providecommand \@ifnum [1]{%
 \ifnum #1\expandafter \@firstoftwo
 \else \expandafter \@secondoftwo
 \fi
}%
\providecommand \@ifx [1]{%
 \ifx #1\expandafter \@firstoftwo
 \else \expandafter \@secondoftwo
 \fi
}%
\providecommand \natexlab [1]{#1}%
\providecommand \enquote  [1]{``#1''}%
\providecommand \bibnamefont  [1]{#1}%
\providecommand \bibfnamefont [1]{#1}%
\providecommand \citenamefont [1]{#1}%
\providecommand \href@noop [0]{\@secondoftwo}%
\providecommand \href [0]{\begingroup \@sanitize@url \@href}%
\providecommand \@href[1]{\@@startlink{#1}\@@href}%
\providecommand \@@href[1]{\endgroup#1\@@endlink}%
\providecommand \@sanitize@url [0]{\catcode `\\12\catcode `\$12\catcode
  `\&12\catcode `\#12\catcode `\^12\catcode `\_12\catcode `\%12\relax}%
\providecommand \@@startlink[1]{}%
\providecommand \@@endlink[0]{}%
\providecommand \url  [0]{\begingroup\@sanitize@url \@url }%
\providecommand \@url [1]{\endgroup\@href {#1}{\urlprefix }}%
\providecommand \urlprefix  [0]{URL }%
\providecommand \Eprint [0]{\href }%
\providecommand \doibase [0]{http://dx.doi.org/}%
\providecommand \selectlanguage [0]{\@gobble}%
\providecommand \bibinfo  [0]{\@secondoftwo}%
\providecommand \bibfield  [0]{\@secondoftwo}%
\providecommand \translation [1]{[#1]}%
\providecommand \BibitemOpen [0]{}%
\providecommand \bibitemStop [0]{}%
\providecommand \bibitemNoStop [0]{.\EOS\space}%
\providecommand \EOS [0]{\spacefactor3000\relax}%
\providecommand \BibitemShut  [1]{\csname bibitem#1\endcsname}%
\let\auto@bib@innerbib\@empty
\bibitem [{\citenamefont {Kane}\ and\ \citenamefont {Mele}(2005)}]{Kane2005}%
  \BibitemOpen
  \bibfield  {author} {\bibinfo {author} {\bibfnamefont {C.~L.}\ \bibnamefont
  {Kane}}\ and\ \bibinfo {author} {\bibfnamefont {E.~J.}\ \bibnamefont
  {Mele}},\ }\href@noop {} {\bibfield  {journal} {\bibinfo  {journal} {Phys.
  Rev. Lett.}\ }\textbf {\bibinfo {volume} {95}},\ \bibinfo {pages} {146802}
  (\bibinfo {year} {2005})}\BibitemShut {NoStop}%
\bibitem [{\citenamefont {Bernevig}\ \emph {et~al.}(2006)\citenamefont
  {Bernevig}, \citenamefont {Hughes},\ and\ \citenamefont
  {Zhang}}]{Bernevig2006}%
  \BibitemOpen
  \bibfield  {author} {\bibinfo {author} {\bibfnamefont {B.~A.}\ \bibnamefont
  {Bernevig}}, \bibinfo {author} {\bibfnamefont {T.~L.}\ \bibnamefont
  {Hughes}}, \ and\ \bibinfo {author} {\bibfnamefont {S.-C.}\ \bibnamefont
  {Zhang}},\ }\href {\doibase 10.1126/science.1133734} {\bibfield  {journal}
  {\bibinfo  {journal} {Science}\ }\textbf {\bibinfo {volume} {314}},\ \bibinfo
  {pages} {1757} (\bibinfo {year} {2006})}\BibitemShut {NoStop}%
\bibitem [{\citenamefont {Koenig}\ \emph {et~al.}(2007)\citenamefont {Koenig},
  \citenamefont {Wiedmann}, \citenamefont {Bruene}, \citenamefont {Roth},
  \citenamefont {Buhmann}, \citenamefont {Molenkamp}, \citenamefont {Qi},\ and\
  \citenamefont {Zhang}}]{Konig2007}%
  \BibitemOpen
  \bibfield  {author} {\bibinfo {author} {\bibfnamefont {M.}~\bibnamefont
  {Koenig}}, \bibinfo {author} {\bibfnamefont {S.}~\bibnamefont {Wiedmann}},
  \bibinfo {author} {\bibfnamefont {C.}~\bibnamefont {Bruene}}, \bibinfo
  {author} {\bibfnamefont {A.}~\bibnamefont {Roth}}, \bibinfo {author}
  {\bibfnamefont {H.}~\bibnamefont {Buhmann}}, \bibinfo {author} {\bibfnamefont
  {L.~W.}\ \bibnamefont {Molenkamp}}, \bibinfo {author} {\bibfnamefont {X.-L.}\
  \bibnamefont {Qi}}, \ and\ \bibinfo {author} {\bibfnamefont {S.-C.}\
  \bibnamefont {Zhang}},\ }\href {\doibase 10.1126/science.1148047} {\bibfield
  {journal} {\bibinfo  {journal} {Science}\ }\textbf {\bibinfo {volume}
  {318}},\ \bibinfo {pages} {766} (\bibinfo {year} {2007})}\BibitemShut
  {NoStop}%
\bibitem [{\citenamefont {Fu}(2011)}]{Fu2011}%
  \BibitemOpen
  \bibfield  {author} {\bibinfo {author} {\bibfnamefont {L.}~\bibnamefont
  {Fu}},\ }\href {\doibase 10.1103/PhysRevLett.106.106802} {\bibfield
  {journal} {\bibinfo  {journal} {Phys. Rev. Lett.}\ }\textbf {\bibinfo
  {volume} {106}},\ \bibinfo {pages} {106802} (\bibinfo {year}
  {2011})}\BibitemShut {NoStop}%
\bibitem [{\citenamefont {Xu}\ \emph {et~al.}(2012)\citenamefont {Xu},
  \citenamefont {Liu}, \citenamefont {Alidoust}, \citenamefont {Neupane},
  \citenamefont {Qian}, \citenamefont {Belopolski}, \citenamefont {Denlinger},
  \citenamefont {Wang}, \citenamefont {Lin}, \citenamefont {Wray},
  \citenamefont {Landolt}, \citenamefont {Slomski}, \citenamefont {Dil},
  \citenamefont {Marcinkova}, \citenamefont {Morosan}, \citenamefont {Gibson},
  \citenamefont {Sankar}, \citenamefont {Chou}, \citenamefont {Cava},
  \citenamefont {Bansil},\ and\ \citenamefont {Hasan}}]{Xu2012}%
  \BibitemOpen
  \bibfield  {author} {\bibinfo {author} {\bibfnamefont {S.-Y.}\ \bibnamefont
  {Xu}}, \bibinfo {author} {\bibfnamefont {C.}~\bibnamefont {Liu}}, \bibinfo
  {author} {\bibfnamefont {N.}~\bibnamefont {Alidoust}}, \bibinfo {author}
  {\bibfnamefont {M.}~\bibnamefont {Neupane}}, \bibinfo {author} {\bibfnamefont
  {D.}~\bibnamefont {Qian}}, \bibinfo {author} {\bibfnamefont {I.}~\bibnamefont
  {Belopolski}}, \bibinfo {author} {\bibfnamefont {J.}~\bibnamefont
  {Denlinger}}, \bibinfo {author} {\bibfnamefont {Y.}~\bibnamefont {Wang}},
  \bibinfo {author} {\bibfnamefont {H.}~\bibnamefont {Lin}}, \bibinfo {author}
  {\bibfnamefont {L.}~\bibnamefont {Wray}}, \bibinfo {author} {\bibfnamefont
  {G.}~\bibnamefont {Landolt}}, \bibinfo {author} {\bibfnamefont
  {B.}~\bibnamefont {Slomski}}, \bibinfo {author} {\bibfnamefont
  {J.}~\bibnamefont {Dil}}, \bibinfo {author} {\bibfnamefont {A.}~\bibnamefont
  {Marcinkova}}, \bibinfo {author} {\bibfnamefont {E.}~\bibnamefont {Morosan}},
  \bibinfo {author} {\bibfnamefont {Q.}~\bibnamefont {Gibson}}, \bibinfo
  {author} {\bibfnamefont {R.}~\bibnamefont {Sankar}}, \bibinfo {author}
  {\bibfnamefont {F.}~\bibnamefont {Chou}}, \bibinfo {author} {\bibfnamefont
  {R.}~\bibnamefont {Cava}}, \bibinfo {author} {\bibfnamefont {A.}~\bibnamefont
  {Bansil}}, \ and\ \bibinfo {author} {\bibfnamefont {M.}~\bibnamefont
  {Hasan}},\ }\href {\doibase 10.1038/ncomms2191} {\bibfield  {journal}
  {\bibinfo  {journal} {Nat. Commun.}\ }\textbf {\bibinfo {volume} {3}},\
  \bibinfo {pages} {1192} (\bibinfo {year} {2012})}\BibitemShut {NoStop}%
\bibitem [{\citenamefont {Dziawa}\ \emph {et~al.}(2012)\citenamefont {Dziawa},
  \citenamefont {Kowalski}, \citenamefont {Dybko}, \citenamefont {Buczko},
  \citenamefont {Szczerbakow}, \citenamefont {Szot}, \citenamefont
  {Lusakowska}, \citenamefont {Balasubramanian}, \citenamefont {Wojek},
  \citenamefont {Berntsen}, \citenamefont {Tjernberg},\ and\ \citenamefont
  {Story}}]{Dziawa2012}%
  \BibitemOpen
  \bibfield  {author} {\bibinfo {author} {\bibfnamefont {P.}~\bibnamefont
  {Dziawa}}, \bibinfo {author} {\bibfnamefont {B.~J.}\ \bibnamefont
  {Kowalski}}, \bibinfo {author} {\bibfnamefont {K.}~\bibnamefont {Dybko}},
  \bibinfo {author} {\bibfnamefont {R.}~\bibnamefont {Buczko}}, \bibinfo
  {author} {\bibfnamefont {A.}~\bibnamefont {Szczerbakow}}, \bibinfo {author}
  {\bibfnamefont {M.}~\bibnamefont {Szot}}, \bibinfo {author} {\bibfnamefont
  {E.}~\bibnamefont {Lusakowska}}, \bibinfo {author} {\bibfnamefont
  {T.}~\bibnamefont {Balasubramanian}}, \bibinfo {author} {\bibfnamefont
  {B.~M.}\ \bibnamefont {Wojek}}, \bibinfo {author} {\bibfnamefont {M.~H.}\
  \bibnamefont {Berntsen}}, \bibinfo {author} {\bibfnamefont {O.}~\bibnamefont
  {Tjernberg}}, \ and\ \bibinfo {author} {\bibfnamefont {T.}~\bibnamefont
  {Story}},\ }\href {\doibase 10.1038/nmat3449} {\bibfield  {journal} {\bibinfo
   {journal} {Nat. Mater.}\ }\textbf {\bibinfo {volume} {11}},\ \bibinfo
  {pages} {1023} (\bibinfo {year} {2012})}\BibitemShut {NoStop}%
\bibitem [{\citenamefont {Hsieh}\ \emph {et~al.}(2012)\citenamefont {Hsieh},
  \citenamefont {Lin}, \citenamefont {Liu}, \citenamefont {Duan}, \citenamefont
  {Bansil},\ and\ \citenamefont {Fu}}]{Hsieh2012}%
  \BibitemOpen
  \bibfield  {author} {\bibinfo {author} {\bibfnamefont {T.~H.}\ \bibnamefont
  {Hsieh}}, \bibinfo {author} {\bibfnamefont {H.}~\bibnamefont {Lin}}, \bibinfo
  {author} {\bibfnamefont {J.}~\bibnamefont {Liu}}, \bibinfo {author}
  {\bibfnamefont {W.}~\bibnamefont {Duan}}, \bibinfo {author} {\bibfnamefont
  {A.}~\bibnamefont {Bansil}}, \ and\ \bibinfo {author} {\bibfnamefont
  {L.}~\bibnamefont {Fu}},\ }\href {\doibase 10.1038/ncomms1969} {\bibfield
  {journal} {\bibinfo  {journal} {Nat. Commun.}\ }\textbf {\bibinfo {volume}
  {3}},\ \bibinfo {pages} {982} (\bibinfo {year} {2012})}\BibitemShut {NoStop}%
\bibitem [{\citenamefont {Tanaka}\ \emph {et~al.}(2012)\citenamefont {Tanaka},
  \citenamefont {Ren}, \citenamefont {Sato}, \citenamefont {Nakayama},
  \citenamefont {Souma}, \citenamefont {Takahashi}, \citenamefont {Segawa},\
  and\ \citenamefont {Ando}}]{Tanaka2012}%
  \BibitemOpen
  \bibfield  {author} {\bibinfo {author} {\bibfnamefont {Y.}~\bibnamefont
  {Tanaka}}, \bibinfo {author} {\bibfnamefont {Z.}~\bibnamefont {Ren}},
  \bibinfo {author} {\bibfnamefont {T.}~\bibnamefont {Sato}}, \bibinfo {author}
  {\bibfnamefont {K.}~\bibnamefont {Nakayama}}, \bibinfo {author}
  {\bibfnamefont {S.}~\bibnamefont {Souma}}, \bibinfo {author} {\bibfnamefont
  {T.}~\bibnamefont {Takahashi}}, \bibinfo {author} {\bibfnamefont
  {K.}~\bibnamefont {Segawa}}, \ and\ \bibinfo {author} {\bibfnamefont
  {Y.}~\bibnamefont {Ando}},\ }\href {\doibase 10.1038/nphys2442} {\bibfield
  {journal} {\bibinfo  {journal} {Nat. Phys.}\ }\textbf {\bibinfo {volume}
  {8}},\ \bibinfo {pages} {800} (\bibinfo {year} {2012})}\BibitemShut {NoStop}%
\bibitem [{\citenamefont {Barone}\ \emph {et~al.}(2013)\citenamefont {Barone},
  \citenamefont {Di~Sante},\ and\ \citenamefont {Picozzi}}]{Barone2013}%
  \BibitemOpen
  \bibfield  {author} {\bibinfo {author} {\bibfnamefont {P.}~\bibnamefont
  {Barone}}, \bibinfo {author} {\bibfnamefont {D.}~\bibnamefont {Di~Sante}}, \
  and\ \bibinfo {author} {\bibfnamefont {S.}~\bibnamefont {Picozzi}},\ }\href
  {\doibase 10.1002/pssr.201308154} {\bibfield  {journal} {\bibinfo  {journal}
  {Physica Status Solidi-R}\ }\textbf {\bibinfo {volume} {7}},\ \bibinfo
  {pages} {1102} (\bibinfo {year} {2013})}\BibitemShut {NoStop}%
\bibitem [{\citenamefont {Zhao}\ \emph {et~al.}(2015)\citenamefont {Zhao},
  \citenamefont {Wang}, \citenamefont {Gu},\ and\ \citenamefont
  {Duan}}]{Zhao2015}%
  \BibitemOpen
  \bibfield  {author} {\bibinfo {author} {\bibfnamefont {L.}~\bibnamefont
  {Zhao}}, \bibinfo {author} {\bibfnamefont {J.}~\bibnamefont {Wang}}, \bibinfo
  {author} {\bibfnamefont {B.-L.}\ \bibnamefont {Gu}}, \ and\ \bibinfo {author}
  {\bibfnamefont {W.}~\bibnamefont {Duan}},\ }\href {\doibase
  10.1103/PhysRevB.91.195320} {\bibfield  {journal} {\bibinfo  {journal} {Phys.
  Rev. B}\ }\textbf {\bibinfo {volume} {91}},\ \bibinfo {pages} {195320}
  (\bibinfo {year} {2015})}\BibitemShut {NoStop}%
\bibitem [{\citenamefont {Sessi}\ \emph {et~al.}(2016)\citenamefont {Sessi},
  \citenamefont {Di~Sante}, \citenamefont {Szczerbakow}, \citenamefont {Glott},
  \citenamefont {Wilfert}, \citenamefont {Schmidt}, \citenamefont {Bathon},
  \citenamefont {Dziawa}, \citenamefont {Greiter}, \citenamefont {Neupert},
  \citenamefont {Sangiovanni}, \citenamefont {Story}, \citenamefont {Thomale},\
  and\ \citenamefont {Bode}}]{Sessi2016}%
  \BibitemOpen
  \bibfield  {author} {\bibinfo {author} {\bibfnamefont {P.}~\bibnamefont
  {Sessi}}, \bibinfo {author} {\bibfnamefont {D.}~\bibnamefont {Di~Sante}},
  \bibinfo {author} {\bibfnamefont {A.}~\bibnamefont {Szczerbakow}}, \bibinfo
  {author} {\bibfnamefont {F.}~\bibnamefont {Glott}}, \bibinfo {author}
  {\bibfnamefont {S.}~\bibnamefont {Wilfert}}, \bibinfo {author} {\bibfnamefont
  {H.}~\bibnamefont {Schmidt}}, \bibinfo {author} {\bibfnamefont
  {T.}~\bibnamefont {Bathon}}, \bibinfo {author} {\bibfnamefont
  {P.}~\bibnamefont {Dziawa}}, \bibinfo {author} {\bibfnamefont
  {M.}~\bibnamefont {Greiter}}, \bibinfo {author} {\bibfnamefont
  {T.}~\bibnamefont {Neupert}}, \bibinfo {author} {\bibfnamefont
  {G.}~\bibnamefont {Sangiovanni}}, \bibinfo {author} {\bibfnamefont
  {T.}~\bibnamefont {Story}}, \bibinfo {author} {\bibfnamefont
  {R.}~\bibnamefont {Thomale}}, \ and\ \bibinfo {author} {\bibfnamefont
  {M.}~\bibnamefont {Bode}},\ }\href {\doibase 10.1126/science.aah6233}
  {\bibfield  {journal} {\bibinfo  {journal} {Science}\ }\textbf {\bibinfo
  {volume} {354}},\ \bibinfo {pages} {1269} (\bibinfo {year}
  {2016})}\BibitemShut {NoStop}%
\bibitem [{\citenamefont {Bauer}(1986)}]{Bauer1986}%
  \BibitemOpen
  \bibfield  {author} {\bibinfo {author} {\bibfnamefont {G.}~\bibnamefont
  {Bauer}},\ }\href {\doibase 10.1016/0039-6028(86)90876-9} {\bibfield
  {journal} {\bibinfo  {journal} {Surf. Sci.}\ }\textbf {\bibinfo {volume}
  {168}},\ \bibinfo {pages} {462 } (\bibinfo {year} {1986})}\BibitemShut
  {NoStop}%
\bibitem [{\citenamefont {Zogg}\ and\ \citenamefont
  {Teodoropol}(1995)}]{Zogg1995}%
  \BibitemOpen
  \bibfield  {author} {\bibinfo {author} {\bibfnamefont {H.}~\bibnamefont
  {Zogg}}\ and\ \bibinfo {author} {\bibfnamefont {S.}~\bibnamefont
  {Teodoropol}},\ }\href {\doibase 10.1016/0022-0248(95)80126-W} {\bibfield
  {journal} {\bibinfo  {journal} {J. Cryst. Growth}\ }\textbf {\bibinfo
  {volume} {150}},\ \bibinfo {pages} {1186 } (\bibinfo {year}
  {1995})}\BibitemShut {NoStop}%
\bibitem [{\citenamefont {Springholz}\ and\ \citenamefont
  {Bauer}(2007)}]{Springholz2007}%
  \BibitemOpen
  \bibfield  {author} {\bibinfo {author} {\bibfnamefont {G.}~\bibnamefont
  {Springholz}}\ and\ \bibinfo {author} {\bibfnamefont {G.}~\bibnamefont
  {Bauer}},\ }\href {\doibase 10.1002/pssb.200675616} {\bibfield  {journal}
  {\bibinfo  {journal} {Physica Status Solidi B}\ }\textbf {\bibinfo {volume}
  {244}},\ \bibinfo {pages} {2752} (\bibinfo {year} {2007})}\BibitemShut
  {NoStop}%
\bibitem [{\citenamefont {Yan}\ \emph {et~al.}(2014)\citenamefont {Yan},
  \citenamefont {Guo}, \citenamefont {Wen}, \citenamefont {Zhang},
  \citenamefont {Wang}, \citenamefont {He}, \citenamefont {Ma}, \citenamefont
  {Ji}, \citenamefont {Chen},\ and\ \citenamefont {Xue}}]{Yan2014}%
  \BibitemOpen
  \bibfield  {author} {\bibinfo {author} {\bibfnamefont {C.-H.}\ \bibnamefont
  {Yan}}, \bibinfo {author} {\bibfnamefont {H.}~\bibnamefont {Guo}}, \bibinfo
  {author} {\bibfnamefont {J.}~\bibnamefont {Wen}}, \bibinfo {author}
  {\bibfnamefont {Z.-D.}\ \bibnamefont {Zhang}}, \bibinfo {author}
  {\bibfnamefont {L.-L.}\ \bibnamefont {Wang}}, \bibinfo {author}
  {\bibfnamefont {K.}~\bibnamefont {He}}, \bibinfo {author} {\bibfnamefont
  {X.-C.}\ \bibnamefont {Ma}}, \bibinfo {author} {\bibfnamefont {S.-H.}\
  \bibnamefont {Ji}}, \bibinfo {author} {\bibfnamefont {X.}~\bibnamefont
  {Chen}}, \ and\ \bibinfo {author} {\bibfnamefont {Q.-K.}\ \bibnamefont
  {Xue}},\ }\href {\doibase 10.1016/j.susc.2013.11.004} {\bibfield  {journal}
  {\bibinfo  {journal} {Surf. Sci.}\ }\textbf {\bibinfo {volume} {621}},\
  \bibinfo {pages} {104} (\bibinfo {year} {2014})}\BibitemShut {NoStop}%
\bibitem [{\citenamefont {Akiyama}\ \emph {et~al.}(2016)\citenamefont
  {Akiyama}, \citenamefont {Fujisawa}, \citenamefont {Yamaguchi}, \citenamefont
  {Ishikawa},\ and\ \citenamefont {Kuroda}}]{Akiyama2016}%
  \BibitemOpen
  \bibfield  {author} {\bibinfo {author} {\bibfnamefont {R.}~\bibnamefont
  {Akiyama}}, \bibinfo {author} {\bibfnamefont {K.}~\bibnamefont {Fujisawa}},
  \bibinfo {author} {\bibfnamefont {T.}~\bibnamefont {Yamaguchi}}, \bibinfo
  {author} {\bibfnamefont {R.}~\bibnamefont {Ishikawa}}, \ and\ \bibinfo
  {author} {\bibfnamefont {S.}~\bibnamefont {Kuroda}},\ }\href {\doibase
  10.1007/s12274-015-0930-8} {\bibfield  {journal} {\bibinfo  {journal} {Nano
  Res.}\ }\textbf {\bibinfo {volume} {9}},\ \bibinfo {pages} {490} (\bibinfo
  {year} {2016})}\BibitemShut {NoStop}%
\bibitem [{\citenamefont {Taskin}\ \emph {et~al.}(2014)\citenamefont {Taskin},
  \citenamefont {Yang}, \citenamefont {Sasaki}, \citenamefont {Segawa},\ and\
  \citenamefont {Ando}}]{Taskin2014}%
  \BibitemOpen
  \bibfield  {author} {\bibinfo {author} {\bibfnamefont {A.~A.}\ \bibnamefont
  {Taskin}}, \bibinfo {author} {\bibfnamefont {F.}~\bibnamefont {Yang}},
  \bibinfo {author} {\bibfnamefont {S.}~\bibnamefont {Sasaki}}, \bibinfo
  {author} {\bibfnamefont {K.}~\bibnamefont {Segawa}}, \ and\ \bibinfo {author}
  {\bibfnamefont {Y.}~\bibnamefont {Ando}},\ }\href {\doibase
  10.1103/PhysRevB.89.121302} {\bibfield  {journal} {\bibinfo  {journal} {Phys.
  Rev. B}\ }\textbf {\bibinfo {volume} {89}},\ \bibinfo {pages} {121302}
  (\bibinfo {year} {2014})}\BibitemShut {NoStop}%
\bibitem [{\citenamefont {Zeljkovic}\ \emph {et~al.}(2015)\citenamefont
  {Zeljkovic}, \citenamefont {Walkup}, \citenamefont {Assaf}, \citenamefont
  {Scipioni}, \citenamefont {Sankar}, \citenamefont {Chou},\ and\ \citenamefont
  {Madhavan}}]{Zeljkovic2015}%
  \BibitemOpen
  \bibfield  {author} {\bibinfo {author} {\bibfnamefont {I.}~\bibnamefont
  {Zeljkovic}}, \bibinfo {author} {\bibfnamefont {D.}~\bibnamefont {Walkup}},
  \bibinfo {author} {\bibfnamefont {B.~A.}\ \bibnamefont {Assaf}}, \bibinfo
  {author} {\bibfnamefont {K.~L.}\ \bibnamefont {Scipioni}}, \bibinfo {author}
  {\bibfnamefont {R.}~\bibnamefont {Sankar}}, \bibinfo {author} {\bibfnamefont
  {F.}~\bibnamefont {Chou}}, \ and\ \bibinfo {author} {\bibfnamefont
  {V.}~\bibnamefont {Madhavan}},\ }\href
  {http://dx.doi.org/10.1038/nnano.2015.177} {\bibfield  {journal} {\bibinfo
  {journal} {Nat. Nanotechnol.}\ }\textbf {\bibinfo {volume} {10}},\ \bibinfo
  {pages} {849} (\bibinfo {year} {2015})}\BibitemShut {NoStop}%
\bibitem [{\citenamefont {Ishikawa}\ \emph {et~al.}(2016)\citenamefont
  {Ishikawa}, \citenamefont {Yamaguchi}, \citenamefont {Ohtaki}, \citenamefont
  {Akiyama},\ and\ \citenamefont {Kuroda}}]{Ishikawa2016}%
  \BibitemOpen
  \bibfield  {author} {\bibinfo {author} {\bibfnamefont {R.}~\bibnamefont
  {Ishikawa}}, \bibinfo {author} {\bibfnamefont {T.}~\bibnamefont {Yamaguchi}},
  \bibinfo {author} {\bibfnamefont {Y.}~\bibnamefont {Ohtaki}}, \bibinfo
  {author} {\bibfnamefont {R.}~\bibnamefont {Akiyama}}, \ and\ \bibinfo
  {author} {\bibfnamefont {S.}~\bibnamefont {Kuroda}},\ }\href {\doibase
  10.1016/j.jcrysgro.2016.08.027} {\bibfield  {journal} {\bibinfo  {journal}
  {J. Cryst. Growth}\ }\textbf {\bibinfo {volume} {453}},\ \bibinfo {pages}
  {124 } (\bibinfo {year} {2016})}\BibitemShut {NoStop}%
\bibitem [{\citenamefont {Tanaka}\ \emph {et~al.}(2013)\citenamefont {Tanaka},
  \citenamefont {Shoman}, \citenamefont {Nakayama}, \citenamefont {Souma},
  \citenamefont {Sato}, \citenamefont {Takahashi}, \citenamefont {Novak},
  \citenamefont {Segawa},\ and\ \citenamefont {Ando}}]{Takana2013}%
  \BibitemOpen
  \bibfield  {author} {\bibinfo {author} {\bibfnamefont {Y.}~\bibnamefont
  {Tanaka}}, \bibinfo {author} {\bibfnamefont {T.}~\bibnamefont {Shoman}},
  \bibinfo {author} {\bibfnamefont {K.}~\bibnamefont {Nakayama}}, \bibinfo
  {author} {\bibfnamefont {S.}~\bibnamefont {Souma}}, \bibinfo {author}
  {\bibfnamefont {T.}~\bibnamefont {Sato}}, \bibinfo {author} {\bibfnamefont
  {T.}~\bibnamefont {Takahashi}}, \bibinfo {author} {\bibfnamefont
  {M.}~\bibnamefont {Novak}}, \bibinfo {author} {\bibfnamefont
  {K.}~\bibnamefont {Segawa}}, \ and\ \bibinfo {author} {\bibfnamefont
  {Y.}~\bibnamefont {Ando}},\ }\href
  {https://link.aps.org/doi/10.1103/PhysRevB.88.235126} {\bibfield  {journal}
  {\bibinfo  {journal} {Phys. Rev. B}\ }\textbf {\bibinfo {volume} {88}},\
  \bibinfo {pages} {235126} (\bibinfo {year} {2013})}\BibitemShut {NoStop}%
\bibitem [{\citenamefont {Springholz}(2003)}]{Khokhlov2002}%
  \BibitemOpen
  \bibfield  {author} {\bibinfo {author} {\bibfnamefont {G.}~\bibnamefont
  {Springholz}},\ }\href {https://books.google.de/books?id=y4P4Kf399l8C} {\emph
  {\bibinfo {title} {Lead Chalcogenides: Physics and Applications}}},\ edited
  by\ \bibinfo {editor} {\bibfnamefont {D.}~\bibnamefont {Khokhlov}}\ (\bibinfo
   {publisher} {Taylor \& Francis},\ \bibinfo {address} {New York},\ \bibinfo
  {year} {2003})\BibitemShut {NoStop}%
\bibitem [{\citenamefont {Grynberg}\ \emph {et~al.}(1996)\citenamefont
  {Grynberg}, \citenamefont {Huant}, \citenamefont {Martinez}, \citenamefont
  {Kossut}, \citenamefont {Wojtowicz}, \citenamefont {Karczewski},
  \citenamefont {Shi}, \citenamefont {Peeters},\ and\ \citenamefont
  {Devreese}}]{Grynberg1996}%
  \BibitemOpen
  \bibfield  {author} {\bibinfo {author} {\bibfnamefont {M.}~\bibnamefont
  {Grynberg}}, \bibinfo {author} {\bibfnamefont {S.}~\bibnamefont {Huant}},
  \bibinfo {author} {\bibfnamefont {G.}~\bibnamefont {Martinez}}, \bibinfo
  {author} {\bibfnamefont {J.}~\bibnamefont {Kossut}}, \bibinfo {author}
  {\bibfnamefont {T.}~\bibnamefont {Wojtowicz}}, \bibinfo {author}
  {\bibfnamefont {G.}~\bibnamefont {Karczewski}}, \bibinfo {author}
  {\bibfnamefont {J.~M.}\ \bibnamefont {Shi}}, \bibinfo {author} {\bibfnamefont
  {F.~M.}\ \bibnamefont {Peeters}}, \ and\ \bibinfo {author} {\bibfnamefont
  {J.~T.}\ \bibnamefont {Devreese}},\ }\href {\doibase
  10.1103/PhysRevB.54.1467} {\bibfield  {journal} {\bibinfo  {journal} {Phys.
  Rev. B}\ }\textbf {\bibinfo {volume} {54}},\ \bibinfo {pages} {1467}
  (\bibinfo {year} {1996})}\BibitemShut {NoStop}%
\bibitem [{\citenamefont {Chen}\ \emph {et~al.}(1995)\citenamefont {Chen},
  \citenamefont {Faurie}, \citenamefont {Sivananthan}, \citenamefont {Hua},\
  and\ \citenamefont {Otsuka}}]{Chen1995}%
  \BibitemOpen
  \bibfield  {author} {\bibinfo {author} {\bibfnamefont {Y.~P.}\ \bibnamefont
  {Chen}}, \bibinfo {author} {\bibfnamefont {J.~P.}\ \bibnamefont {Faurie}},
  \bibinfo {author} {\bibfnamefont {S.}~\bibnamefont {Sivananthan}}, \bibinfo
  {author} {\bibfnamefont {G.~C.}\ \bibnamefont {Hua}}, \ and\ \bibinfo
  {author} {\bibfnamefont {N.}~\bibnamefont {Otsuka}},\ }\href {\doibase
  10.1007/BF02657950} {\bibfield  {journal} {\bibinfo  {journal} {J. Electron.
  Mater.}\ }\textbf {\bibinfo {volume} {24}},\ \bibinfo {pages} {475} (\bibinfo
  {year} {1995})}\BibitemShut {NoStop}%
\bibitem [{\citenamefont {Tarakina}\ \emph {et~al.}()\citenamefont {Tarakina},
  \citenamefont {Schreyeck}, \citenamefont {Luysberg}, \citenamefont {Grauer},
  \citenamefont {Schumacher}, \citenamefont {Karczewski}, \citenamefont
  {Brunner}, \citenamefont {Gould}, \citenamefont {Buhmann}, \citenamefont
  {Dunin-Borkowski},\ and\ \citenamefont {Molenkamp}}]{tarakina2014}%
  \BibitemOpen
  \bibfield  {author} {\bibinfo {author} {\bibfnamefont {N.~V.}\ \bibnamefont
  {Tarakina}}, \bibinfo {author} {\bibfnamefont {S.}~\bibnamefont {Schreyeck}},
  \bibinfo {author} {\bibfnamefont {M.}~\bibnamefont {Luysberg}}, \bibinfo
  {author} {\bibfnamefont {S.}~\bibnamefont {Grauer}}, \bibinfo {author}
  {\bibfnamefont {C.}~\bibnamefont {Schumacher}}, \bibinfo {author}
  {\bibfnamefont {G.}~\bibnamefont {Karczewski}}, \bibinfo {author}
  {\bibfnamefont {K.}~\bibnamefont {Brunner}}, \bibinfo {author} {\bibfnamefont
  {C.}~\bibnamefont {Gould}}, \bibinfo {author} {\bibfnamefont
  {H.}~\bibnamefont {Buhmann}}, \bibinfo {author} {\bibfnamefont {R.~E.}\
  \bibnamefont {Dunin-Borkowski}}, \ and\ \bibinfo {author} {\bibfnamefont
  {L.~W.}\ \bibnamefont {Molenkamp}},\ }\href {\doibase 10.1002/admi.201400134}
  {\bibfield  {journal} {\bibinfo  {journal} {Advanced Materials Interfaces}\
  }\textbf {\bibinfo {volume} {1}},\ \bibinfo {pages} {1400134}}\BibitemShut
  {NoStop}%
\bibitem [{\citenamefont {Wang}\ \emph {et~al.}(2014)\citenamefont {Wang},
  \citenamefont {Liu}, \citenamefont {Xu}, \citenamefont {Wu}, \citenamefont
  {Gu},\ and\ \citenamefont {Duan}}]{Wang2014}%
  \BibitemOpen
  \bibfield  {author} {\bibinfo {author} {\bibfnamefont {J.}~\bibnamefont
  {Wang}}, \bibinfo {author} {\bibfnamefont {J.}~\bibnamefont {Liu}}, \bibinfo
  {author} {\bibfnamefont {Y.}~\bibnamefont {Xu}}, \bibinfo {author}
  {\bibfnamefont {J.}~\bibnamefont {Wu}}, \bibinfo {author} {\bibfnamefont
  {B.-L.}\ \bibnamefont {Gu}}, \ and\ \bibinfo {author} {\bibfnamefont
  {W.}~\bibnamefont {Duan}},\ }\href {\doibase 10.1103/PhysRevB.89.125308}
  {\bibfield  {journal} {\bibinfo  {journal} {Phys. Rev. B}\ }\textbf {\bibinfo
  {volume} {89}},\ \bibinfo {pages} {125308} (\bibinfo {year}
  {2014})}\BibitemShut {NoStop}%
\bibitem [{\citenamefont {Jiang}\ \emph {et~al.}(2012)\citenamefont {Jiang},
  \citenamefont {Sun}, \citenamefont {Chen}, \citenamefont {Wang},
  \citenamefont {Li}, \citenamefont {Song}, \citenamefont {He}, \citenamefont
  {Wang}, \citenamefont {Chen}, \citenamefont {Xue}, \citenamefont {Ma},\ and\
  \citenamefont {Zhang}}]{Jiang2012}%
  \BibitemOpen
  \bibfield  {author} {\bibinfo {author} {\bibfnamefont {Y.}~\bibnamefont
  {Jiang}}, \bibinfo {author} {\bibfnamefont {Y.~Y.}\ \bibnamefont {Sun}},
  \bibinfo {author} {\bibfnamefont {M.}~\bibnamefont {Chen}}, \bibinfo {author}
  {\bibfnamefont {Y.}~\bibnamefont {Wang}}, \bibinfo {author} {\bibfnamefont
  {Z.}~\bibnamefont {Li}}, \bibinfo {author} {\bibfnamefont {C.}~\bibnamefont
  {Song}}, \bibinfo {author} {\bibfnamefont {K.}~\bibnamefont {He}}, \bibinfo
  {author} {\bibfnamefont {L.}~\bibnamefont {Wang}}, \bibinfo {author}
  {\bibfnamefont {X.}~\bibnamefont {Chen}}, \bibinfo {author} {\bibfnamefont
  {Q.-K.}\ \bibnamefont {Xue}}, \bibinfo {author} {\bibfnamefont
  {X.}~\bibnamefont {Ma}}, \ and\ \bibinfo {author} {\bibfnamefont {S.~B.}\
  \bibnamefont {Zhang}},\ }\href {\doibase 10.1103/PhysRevLett.108.066809}
  {\bibfield  {journal} {\bibinfo  {journal} {Phys. Rev. Lett.}\ }\textbf
  {\bibinfo {volume} {108}},\ \bibinfo {pages} {066809} (\bibinfo {year}
  {2012})}\BibitemShut {NoStop}%
\bibitem [{\citenamefont {Springholz}\ and\ \citenamefont
  {Wiesauer}(2001)}]{Springholz2001}%
  \BibitemOpen
  \bibfield  {author} {\bibinfo {author} {\bibfnamefont {G.}~\bibnamefont
  {Springholz}}\ and\ \bibinfo {author} {\bibfnamefont {K.}~\bibnamefont
  {Wiesauer}},\ }\href {\doibase 10.1103/PhysRevLett.88.015507} {\bibfield
  {journal} {\bibinfo  {journal} {Phys. Rev. Lett.}\ }\textbf {\bibinfo
  {volume} {88}},\ \bibinfo {pages} {015507} (\bibinfo {year}
  {2001})}\BibitemShut {NoStop}%
\bibitem [{\citenamefont {Knox}\ \emph {et~al.}(2014)\citenamefont {Knox},
  \citenamefont {Bozin}, \citenamefont {Malliakas}, \citenamefont
  {Kanatzidis},\ and\ \citenamefont {Billinge}}]{Knox2014}%
  \BibitemOpen
  \bibfield  {author} {\bibinfo {author} {\bibfnamefont {K.~R.}\ \bibnamefont
  {Knox}}, \bibinfo {author} {\bibfnamefont {E.~S.}\ \bibnamefont {Bozin}},
  \bibinfo {author} {\bibfnamefont {C.~D.}\ \bibnamefont {Malliakas}}, \bibinfo
  {author} {\bibfnamefont {M.~G.}\ \bibnamefont {Kanatzidis}}, \ and\ \bibinfo
  {author} {\bibfnamefont {S.~J.~L.}\ \bibnamefont {Billinge}},\ }\href
  {\doibase 10.1103/PhysRevB.89.014102} {\bibfield  {journal} {\bibinfo
  {journal} {Phys. Rev. B}\ }\textbf {\bibinfo {volume} {89}},\ \bibinfo
  {pages} {014102} (\bibinfo {year} {2014})}\BibitemShut {NoStop}%
\end{thebibliography}

%

\end{document}